\documentstyle[prd,aps,epsfig,preprint]{revtex}
\begin{document}
\draft

\title{\bf Reply to Comment on Conservative Force Fields in Nonextensive
Kinetic Theory}

\author{J.A.S. Lima$^{1}$\footnote {limajas@dfte.ufrn.br},
J.R.Bezerra$^{1}$\footnote{zero@dfte.ufrn.br},
R.Silva$^{2}$\footnote {rsilva@uern.br}}

\smallskip
\address{~\\$^{1}$Universidade
Federal do Rio Grande do Norte,
\\Departamento de F\'{\i}sica,
Caixa Postal 1641, \\59072-970 Natal, RN, Brazil}

\address{~\\$^{2}$Universidade do Estado do Rio Grande do Norte,
\\Departamento de F\'{\i}sica,
\\59610-210 Mossor\'o, RN, Brazil}

\date{\today}
\maketitle
\begin{abstract}
It is shown that the comment by Costa and Meneses\cite{CM03} did
not point any technical or conceptual flaw in our paper titled
``Conservative Force Fields in Nonextensive Kinetic Theory"
(Physica {\bf A 316}, 289 (2002)). In particular, the application
of the nonextensive distribution to the historical problem of the
unbounded isothermal atmosphere under constant gravity is
theoretically correct. It should be stressed that our solution is
somewhat connected with a similar solution to a problem appearing
in the astrophysical domain, namely, the so-called singular
isothermal sphere. Actually, both problems are solved by a
stationary Tsallis q-distribution with a ``thermal cut-off" in the
allowed values of the energy.
\end{abstract}

\newpage
In a recent comment, da Costa and de Meneses[1] raised some
objections concerning the application of the nonextensive kinetic
distribution to the problem of the isothermal atmosphere under
constant gravity as worked out by Lima et al.\cite{LBS}. Ab
initio, we notice that the leitmotiv of our paper was to
investigate how the molecular motion as predicted by the
nonextensive kinetic theory (based on Tsallis
statistics\cite{CT}), is modified by an external force field.
Since this part of the paper (the largest and most important one)
has not been contested in the comment, it is very hard to
understand the logic behind the objection to the most elementary
application - which has been considered mainly to exemplify the
usefulness our approach. In spite of that, we will try to expose
more pedagogically our viewpoint concerning the modified
barometric formula which is a direct consequence of equation (20)
in our paper\cite{LBS}. Hopefully, the comments below may clarify
the basic results of our paper, as well as their physical
interest.

The problem of the isothermal atmosphere under constant gravity
has kinetically been described with basis on the Maxwell-Boltzmann
(MB) distribution, the main prediction of which is the so-called
barometric formula

\begin{equation}\label{e1}
\rho(z)= \rho_0\exp\left[{-\frac{mgz}{k_BT}}\right] \quad.
\end{equation}
The first interesting conclusion from the above expression is that
the ideal isothermal atmosphere in the MB approach extends to an
infinite distance. In their comment, the authors stressed that $g$
and $T$ are constant as $z$ take values on the interval
[$0,\infty$]. They forgot, however, that the infinite variation of
$z$ has been established a posteriori, that is, after the complete
solution of the problem. Indeed, this uncomfortable result was
discussed long ago by Lord Rayleigh\cite{LR}, who attempted to
solve it by replacing the isothermal equilibrium by an adiabatic
condition, thereby getting a finite length to the atmosphere under
constant gravity. In this connection, we notice that whether the
isothermal MB approach describes or not, the observed earth's
atmosphere is not the important point here since the earth's
atmosphere is a very complex system. Its gravitational field is
not constant and the earth's curvature also needs to be taken into
account in order to have a more realistic description. Probably,
more important still, it is a system very far from equilibrium
since turbulence is usually present in many relevant scales.

In this way, the main aspect to reflect in our paper is that the
standard Maxwell-Boltzmann kinetic theory in the presence of
gravity routinely lead to distributions with infinite scales of
length and mass, as happens to the case of the singular isothermal
sphere in the astrophysical domain[5-8]. In the MB approach, such
distributions need to be somewhat truncated in order to furnish a
more reasonable (finite) description, as in the case of stellar
systems. Therefore, although considering that discussions of the
barometric formula are available in any textbook for undergraduate
students (as pointed out by da Costa and de Meneses), we believe
with basis on the above comments that they did not catch the
important points we are talking about. More precisely, MB
distribution in the presence of gravity usually predicts unbounded
systems like the singular isothermal sphere (where ${\vec g}$ is
not constant) because the particles moving in the ``hot tail" of
the distribution always escape from the majority of the physical
potentials satisfying simultaneously the Vlasov and Poisson
equations[4-8]. This explains the interest for the q-nonextensive
class of distributions with a ``thermal cut-off" since the
velocity of the particles (or more generally the total energy) has
an upper limit.

Therefore, unlike the claims of da Costa and de Meneses, our
rediscussion of the barometric formula seems to be relevant to the
enlarged nonextensive framework. They also remarked that any
discussion related to the isothermal atmosphere under constant
gravity must start from the relevant scale of the problem, namely
$\xi = k_BT/mg$. At this point it is interesting to see how such
scale appears in our treatment.

By separating the variables in the Vlasov equation we have shown
that the potential energy term can rigorously be introduced in the
q-nonextensive velocity distribution. The main result is (see
Eq.(20) in \cite{LBS})

\begin{equation}\label{e2}
f({\bf r},v) = B_q\left[1 -(1-q)\left(\frac{m{\bf v}^2}{2k_BT} +
\frac{U({\bf r})}{k_BT}\right)\right]^{1/(1-q)}
\end{equation}
Now, for a constant gravitational field, $U(z)= mgz$, one has
\begin{equation}\label{e3}
f({\bf r},v) = B_q\left[1 -(1-q)\left( \frac{m{\bf v}^2}{2k_BT} +
\frac{mgz}{k_BT}\right)\right]^{1/(1-q)} \quad,
\end{equation}
and by integrating in the velocity space we have the modified
barometric formula:
\begin{equation}
\label{e4} \rho(z) = \rho_0\left[1
-(1-q)\frac{mgz}{k_BT}\right]^{(5-3q)/2(1-q)}.
\end{equation}
Note the presence of the required length scale $\xi$. However, for
a given value of $q<1$, a new relevant macroscopic scale is
introduced by the ``thermal cut-off" of the distribution function
(note that this does not happen for $q\geq 1$). In fact, it is
also a function of $\xi$, however, it determines the extension of
the atmosphere to be
\begin{equation}\label{e5}
z_{max} = {k_{B}T \over mg(1-q)} \quad.
\end{equation}
As remarked before, the above equation is a trivial consequence of
equation (20) in our paper, and therefore, of the q-nonextensive
approach which has not been questioned by the authors.

Concerning the estimates presented in Table 1 of our paper (based
in the above formula), the authors suggest that we would consider
gases more abundant in the atmosphere ($N_2, CO_2$..etc.) and
perhaps more appropriate values of the nonextensive parameter in
order to have better results (to be true, at this point the focus
of the criticism seems to have somewhat shifted). Nevertheless, we
have avoided to consider a long list of gases and many different
values of $q$ because we focus just on the basic result, that is,
the fact that the isothermal atmosphere becomes finite for $q<1$.
Therefore, apart the misprint (the length of the atmosphere in the
oxygen case is 38.7 Km instead of 33.7 Km), the results presented
in table 1 are significant in order to understand the combined
effects of the $q$ parameter and the mass of the particles.

Finally, we stress that the discussion of any gravitational system
with basis on the stationary power law nonextensive distribution
is quite interesting nowadays. Actually, we believe that the
introduction of finite results either for the isothermal
atmosphere under constant gravity, or more generally for globular
clusters and elliptical galaxies, do not represent only a nice
coincidence. Unlike the claims of da Costa and de Meneses, such
results are signalizing that nonextensive effects are relevant in
the astrophysical domain, and, presumably, this kind of research
will be intensified in the near future.

\vspace{0.5cm}

\noindent {\bf Acknowledgments:} This work was supported by
Pronex/FINEP (No. 41.96.0908.00), Conselho Nacional de
Desenvolvimento Cient\'{\i}fico e Tecnol\'{o}gico - CNPq and CAPES
(Brazilian Research Agencies).

\end{document}